\documentclass[11pt]{article}
\usepackage{hyperref}
\pdfoutput=1
\begin{document}
\title{A Device to Measure the Propulsive Power of Nematodes}
\author{Jinzhou Yuan$^1$, Han-Sheng Chuang$^3$, Michael Gnatt$^4$, \\ David M. Raizen$^2$ and Haim H. Bau$^1$ \\
\\\vspace{6pt} $^1$Mechanical Engineering and Applied Mechanics, \\ $^2$Department of Neurology, \\ University of Pennsylvania, Philadelphia, PA 19104, USA \\ $^3$Department of Biomedical Engineering, \\ National Cheng Kung University, Tainan, Taiwan \\ $^4$Mechanical Engineering, \\ University of Maryland, Baltimore County, MD 21250, USA}
\maketitle
\begin{abstract}
The nematodes {\it Caenorhabditis (C.) elegans} is often used as model biological system to study the mechanics of swimming, genetic basis of behavior, disease-progression, and aging, as well as to develop new therapies and screen drugs. On occasion, it is desirable to quantify the nematode’s muscle power as a function of mutation, age, food supply, disease progression, drugs, and other environmental conditions. In the fluid dynamics video, we present a microfluidic device to measure the propulsive power of nematodes. The device consists of a tapered conduit filled with aqueous solution. The conduit is subjected to a DC electric field with the negative pole at the narrow end and to pressure-driven flow directed from the narrow end. The nematode is inserted at the conduit’s wide end. Directed by the electric field (through electrotaxis \cite{1} ), the nematode swims deliberately upstream toward the negative pole of the DC field. As the conduit narrows, the average fluid velocity and the drag force on the nematode increase. Eventually, the nematode arrives at an equilibrium position, at which its propulsive force balances the viscous drag force induced by the adverse flow. Swimming gait, amplitude, and frequency of worms as functions of position along the conduit were obtained using a custom Matlab$^{TM}$ program that we have developed. The swimming amplitude and frequency were found to be nearly independent of the width of the conduit. The equilibrium position of different animals, with similar body lengths, was measured as a function of the flow rate. The flow field around the nematode was obtained by direct numerical simulations with the experimentally imaged gait and the tapered geometry of the conduit as boundary conditions. The flow field generated by a swimming worm is similar to the one induced by two pairs of counter rotating rotors. Equilibrium positions under different flow rates were identified by finding the positions at which the horizontal component of the total force exerted on the worm body vanishes. The theoretically predicted equilibrium positions were compared and favorably agreed with the experimental data. The nematode’s propulsive power was calculated by integrating the product of velocity and total stress over the worm’s body surface. The device is useful to retain the nematodes at a nearly fixed position for prolonged observations of active animals under a microscope, to keep the nematode exercising, and to estimate the nematode’s power consumption based on the conduit’s width at the equilibrium position.
\end{abstract}

\end{document}